\documentstyle[12pt]{article}
\date{}
\title{Generalized bit-moments and cumulants   \\
 based on discrete derivative}
\author{{Ramandeep S. Johal}\\
{\it Institut f\"{u}r Theoretische Physik,}\\
{\it Technische Universit\"{a}t Dresden, }\\
{\it  01062 Dresden, Germany.}\\
e-mail: rjohal$@$theory.phy.tu-dresden.de\\
Ph.:  + 49 (0351) 463 35582  \\
Fax:  + 49 (0351) 463 37299}
\begin{document}
\maketitle
\def\be{\begin{equation}}
\def\ee{\end{equation}}
\def\ba{\begin{eqnarray}}
\def\ea{\end{eqnarray}}
\baselineskip 24pt
\begin{abstract}
We give a simple recipe based on the use of discrete derivative,
to obtain generalized bit-moments obeying 
nonadditive statistics of Tsallis. The generalized
 bit-cumulants may be of two kinds, first which 
preserve the standard relations between moments
and cumulants of a distribution, and are nonadditive
with respect to independent subsystems. The  second 
kind do not preserve usual moment-cumulant relations.
These are additive in nature and Renyi entropy is 
naturally incorporated as the cumulant of order one. 
\end{abstract}


   
\newpage
\section*{1. Introduction}
The last decade saw an increasing interest in a nonextensive
generalization of Boltzmann-Gibbs statistical mechanics, popularly
called as Tsallis statistics. It is based on a nonadditive 
entropy, dependent on a real positive parameter $q$.
Maximising Tsallis entropy under suitable constraints,
 leads to power-law type distributions which have
been applied to various classes of physical systems
and models, for example, fully developed turbulence,
low-dimensional maps at onset of chaos, 
Hamiltonian systems with long-range
interactions etc. (see Ref. \cite{cbpf} for a complete list
of references.) Apart from these, it has motivated
extension of the axiomatic foundations of information theory
\cite{it} as well as development of new mathematical tools 
and quantifiers \cite{qmat}.

In this spirit, the classical bit cumulants were generalized \cite{rrrj}       
 within nonextensive thermodynamic approach.
Bit statistics is a tool to describe the complicated
probability distributions, such as generated by
chaotic systems. Particularly, the second bit cumulant
which is a generalization of specific heat, can be applied in
the context of equilibrium and non-equilibrium phase transitions
\cite{shl}.
The $q$-generalized second bit cumulant has been  applied \cite{tirn} 
 to study the issue of  sensitivity to initial 
conditions in asymmetric unimodal maps
  at the threshold of chaos.         

In this letter, we show that these generalized 
bit- moments and cumulants can be generated simply 
by introducing  a discrete derivative in place of usual
derivative into their 
definitions.
The generalization we aim at can be achieved in two ways:
 (i) only generalize the bit-moments and then use the standard
relations between bit-moments and bit-cumulants to define
generalized cumulants, which yields $q$-cumulants of Ref. \cite{rrrj};
 (ii) generalize the moments and cumulants independently
from their basic definition (see below), using the 
discrete derivative. This procedure 
does not preserve standard relations between moments and cumulants.

First we briefly review the  moments and cumulants in the
Shannon-Gibbs bit statistics.
Given a probability distribution  $\{p_i|x=x_i\}_{i=1,...,W}$
for a discrete random variable $x$, $k$th order moments  
 of the distribution are defined as
$M_k = \sum_{i=1}^{W}{x_i}^k p_i$. It is useful
to define a generalized partition function,
$Z(\sigma) = \sum_{i}p_i e^{\sigma x_i}$.
Then moments  of the distribution $\{p_i\}$ are given as
\be
{M_k = {\partial^k \over \partial\sigma^k}Z(\sigma)
 }\bigg{\vert}_{\sigma =0}.
\label{dmk}
\ee
Similarly, the cumulants for the distribution
may be obtained from a generating function 
$G(\sigma) = {\rm ln}\;Z$, as follows
\be
C_k = {\partial^k \over \partial\sigma^k}G(\sigma)
\bigg{\vert}_{\sigma =0}.
\label{dck}
\ee
 Let us take $x$
to be the fluctuating bit number, $-{\rm ln}\;p_i \equiv -a_i$.
Thus the relevant partition function is
\be
Z(\sigma) = \sum_{i}p_i e^{-\sigma a_i}.
\label{pf}
\ee
Now within the above statistical analysis,
the first cumulant $C_1$ (which is also equal to first moment $M_1$)
is the average value of the bit number.
In the present context, it is  equal to Shannon 
entropy.
The second cumulant is given by $C_2 = \sum_{i}({\rm ln}\;p_i)^2 p_i 
-  (\sum_i p_i{\rm ln}\;p_i)^2$, and is called the bit variance.   
Realizing that within Tsallis' framework, the  generalized
bit number can be  given by 
\be
 -[a_i] = {(p_i)^{-\Delta}-1\over {\Delta}},
\label{gb}
\ee
where from now on we imply $\Delta= (1-q)$,
a generalization \cite{rrrj} of standard cumulants 
was  considered by modifying the partition
function in Eq. (\ref{pf}) based on $[a_i]$
instead of  $a_i$. Such a generalization of 
second bit cumulant has found an interesting
application in low-dimensional maps at chaotic threshold.
 Note that in  applications 
of Tsallis formalism to such systems 
 a critical value of $q=q^*$ exists, which 
  may be inferred from the study
of sensitivity to initial condition \cite{costa} or from the
multifractal spectrum of the attractor at 
chaotic threshold \cite{tslyr}. 
Although, a priori, $q$ introduced above
is a free parameter, it is found to behave like the critical
 index $q^*$ in quantifiers such as  the 
generalized second bit cumulant. This implies that \cite{tirn}
information about relation of Tsallis index $q^*$
and inflexion parameter $z$ of the nonlinear mapping
can be reliably obtained, where it is difficult to
do so by other techniques (see also \cite{tirn2}).
\subsection*{2. Method (i)}
In the following, we show that  moments of the above bit 
number in Eq. (\ref{gb}) 
 can  be obtained by modifying the
derivative rule in the Eq. (\ref{dmk}),  
but keeping the partition function of Eq. (\ref{pf})
unchanged.
We simply replace the ordinary derivative with
respect to $\sigma$ by a discrete derivative 
operator defined as follows
\be
\partial_{{\Delta},x}f(x) = {f(x+\Delta) - f(x)\over \Delta}.
\label{dd}
\ee
Naturally, as $\Delta=(1-q)\to 0^+$, we get back the ordinary derivative.
Now we make the substitution
\be
{\partial^k \over \partial\sigma^k}Z(\sigma)\to 
{\partial^{k}_{{\Delta},\sigma}}Z(\sigma).
\label{subs}
\ee
In other words,
 the $q$-moments of 
order $k$, are  defined as 
\be
M^{(q)}_{k} = {\partial^{k}_{{\Delta},\sigma}}Z(\sigma)
{\vert}_{\sigma=0}.
\label{mkq}
\ee
It can be easily verified that the $q$-moment of order 1
 is  
\be
M^{(q)}_{1} = {1-\sum_{i} (p_i)^q \over q-1},
\ee
which is Tsallis entropy. Pseudoadditivity of Tsallis entropy
can be understood as originating from the modified Leibnitz rule
\ba
\partial_{\Delta,x}fg(x) &=& \partial_{\Delta,x}f(x)g(x)
                            + f(x+\Delta)\partial_{\Delta,x}g(x)\\
                         &=& \partial_{\Delta,x}f(x)g(x +\Delta)
                            + f(x)\partial_{\Delta,x}g(x).
\ea
In general, we have 
\be
M^{(q)}_{k} = \sum_i (-[a_i])^k p_i.
\ee
Note that generalized bit number $-[a_i]$  of Eq. (\ref{gb})
appears automatically by operating discrete derivative 
on the "classical" partition function, Eq. (\ref{pf}).
To obtain $q$-cumulants, we have to consider
the rule in Eq. (\ref{subs}) again. For
that purpose we note that
\be
C_1 = {1\over Z} {\partial \over \partial \sigma}
        Z(\sigma)\bigg{\vert}_{\sigma=0}.
\label{c1}
\ee
Thus $q$-cumulant of order 1 can be written as
\be
C^{(q)}_{1} = {1\over Z} {\partial_{{\Delta},\sigma}}Z(\sigma)
\bigg{\vert}_{\sigma=0}.
\ee
Due to the normalization $Z(\sigma=0) = 1$, $C^{(q)}_{1}=M^{(q)}_{1}$.
Similarly, we have
\be
C_2 = {1\over Z}{\partial^2 \over \partial\sigma^2}Z(\sigma)
\bigg{\vert}_{\sigma=0} - {1\over Z^2}\left({\partial Z(\sigma) \over \partial 
\sigma}\right)^2\bigg{\vert}_{\sigma=0}.
\label{c2}
\ee
Applying rule of Eq. (\ref{subs}), we have 
\be
C^{(q)}_{2} = {1\over Z}{\partial^{2}_{{\Delta},\sigma}}Z(\sigma)
\bigg{\vert}_{\sigma=0} -
              {1\over Z^2}({\partial_{{\Delta},\sigma}}Z(\sigma)
    )^2\bigg{\vert}_{\sigma=0},
\label{c2q}
\ee
and so on for higher order $q$-cumulants.
Note that $C^{(q)}_{2}$ is the variance of the generalized bit number
$-[a_i]$ in the same way as $C_2$ is the variance of the classical
bit number $-{\rm ln}\;p_i$. 
In the above scheme, we have effectively generalized the 
standard bit moments to $q$-moments and then 
$q$-cumulants follow as a result of standard 
relations between moments and cumulants
of a distribution. In this way, we maintain the usual meaning
of moments and cumulants of a distribution. 
\subsection*{3. Method (ii)}
Naturally, one cannot claim a  unique way to define the generalized bit 
moments as well as the cumulants. This problem is similar
to the one of defining a (multi)parameter dependent generalization of 
Shannon entropy, where some 
physical/mathematical principle is required 
to motivate the generalization. For example,   
 seeking the  form invariance of certain
mathematical relations can require  the valid generalized entropy to be 
an appropriate renormalisation of
Tsallis entropy \cite{abraj}. In this sense,
 our method (i) is motivated by the requirement 
of keeping the standard meaning of bit moments and cumulants,  
and also the relations between them.
Otherwise, the use of a modified  derivative operator
offers at least one more possibility to define a 
generalization of the standard bit cumulants.
The new rule may  be given as  
\be
{\partial^k \over \partial\sigma^k}\to
{\partial^{k}_{{\Delta},\sigma}}.
\label{subs2}
\ee
Compare this with Eq. (\ref{subs}).
Then applying the above rule to Eqs. (\ref{dmk}) and (\ref{dck}),
we note that $q$-moments remain the same as given by
Eq. (\ref{mkq}), but $q$-cumulants are a new set of 
quantities given by
\be
{\cal C}^{(q)}_{k} = \partial_{{\Delta},\sigma}^{k} 
 G(\sigma){\vert}_{\sigma =0}.
\label{dck2}
\ee
This generalization seems interesting  due to the fact
that the first $q$-cumulant is 
\be
{\cal C}^{(q)}_{1} = {{\rm ln}\;\sum_{i} (p_i)^{q} \over 1-q},
\ee
which is Renyi entropy with index $q$ \cite{reni}.
{\it This distinguishes Tsallis entropy and Renyi entropy,
based on the use of discrete derivative: Tsallis
entropy is obtained from $Z$, while Renyi entropy follows
from the generating function $G={\rm ln}\;Z$}.
Also this definition of Renyi entropy potentially connects
 it to non-commutative differential calculus \cite{dima}
 which  is based on discrete derivative.

However, it should be pointed out that in method (ii), the standard
relations between moments and cumulants cannot
be preserved. Thus as is already evident, the
first $q$-cumulant is not equal to first $q$-moment.
This implies that  Renyi entropy cannot be inferred here as 
a simple expectation value of some generalized bit number.
In fact, it can be written only as a kind of nonlinear average
 \cite{kn}. 

Similarly, the $q$-cumulant of order 2 is given by
\be
{\cal C}^{(q)}_{2} = {1\over (1-q)^2}\left[{\rm ln}\;\sum_{i} (p_i)^{2q-1}
                    - 2{\rm ln}\;\sum_{i} (p_i)^{q} \right],
\label{c22}
\ee
which is not equal to the bit variance of generalized bit
number $-[a_i]$, given from Eq. (\ref{c2q}) as 
\be
C^{(q)}_{2} = {1\over (1-q)^2}\left[\sum_{i} (p_i)^{2q-1}-
                   (\sum_{i} (p_i)^{q})^2\right].
\ee
Thus {\it there does not appear to be a generalized
bit number connected with Renyi entropy, unlike $-[a_i]$ for
Tsallis entropy.} 
However, these quantities do  go back to the standard bit variance
 in the limit $q\to 1$. The positivity of ${\cal C}^{(q)}_{2}$ follows from the 
similar property of $C^{(q)}_{2}$. Thus if ${\cal C}^{(q)}_{2}$
is considered as generalization of $C^{(1)}_{2}$, then 
 it is interesting to see  the implications of 
using ${\cal C}^{(q)}_{2}$ instead of $C^{(q)}_{2}$ in such systems
as, for example, low dimensional maps along the lines of Ref. \cite{tirn}.
Such a comparative study can help to elucidate the separate
roles of Renyi and Tsallis entropies in these systems when they exhibit 
power-law behaviour \cite{rjut}.

After observing that discrete derivative may be   
relevant for Renyi entropy, 
we consider the special case of multifractals \cite{feder}.
For such measures also, we can define a
generalized partition function $Z(\beta) = \sum_i (p_i)^{\beta}$,
where the sum ranges over boxes of size $l$.
We note that $Z(\beta)$ and $Z(\sigma)$ in Eq. (\ref{pf})
are equivalent and $\beta \equiv 1-\sigma$.
Thus discrete derivative can be expected to be significant
for multifractal measures also. 
For length scale $l\to 0$, the partition function scales
as
\be Z(\beta)= \sum_i (p_i)^{\beta}\sim l^{-\tau(\beta)}.
\ee
The measure is characterized by a whole sequence
of exponents $\tau(\beta)$, given by
\be
\tau(\beta) = -\lim_{l\to 0} { {\rm ln}\; \sum_i (p_i)^{\beta}
               \over {\rm ln}\;l}.
\ee
Choosing different values of $\beta$ helps to scan subsets with
different probability strengths scaling as $p_i \sim l^{\alpha_i}$.
These properties can be expressed by taking the derivative
of $\tau(\beta)$ as
\be
{d\over d\beta}\tau(\beta) = -\lim_{l\to 0}{\sum_i (p_i)^{\beta}
 {\rm ln}\;p_i \over \sum_i (p_i)^{\beta} {\rm ln}\;l}.
\label{dtb}
\ee
Let us replace the ordinary derivative by the discrete derivative
and taking into the fact that $\beta=1-\sigma$, we find that
\be
\partial_{{\Delta},\beta} \tau(\beta) =
-\lim_{l\to 0}\; {\rm ln}\left({\sum_i (p_i)^{\beta -\Delta}
\over \sum_i (p_i)^{\beta}}\right) \bigg{/} \Delta{\rm ln}\;l.
\ee
We find that
\be
\partial_{{\Delta},\beta} \tau(\beta)\vert_{\beta\to \infty}
  = -\alpha_{\rm max},
\ee
and
\be
\partial_{{\Delta},\beta} \tau(\beta)\vert_{\beta\to -\infty}
 =-\alpha_{\rm min},
\ee
where
$\alpha_{\rm max}$ and $\alpha_{\rm min}$ are the
end points of the multifractal spectrum.
These results are  just the same as will be obtained by working with
Eq. (\ref{dtb}).
This is reasonable, as discrete derivative in $\beta$ measures
the change in function with respect to a finite increment
$\Delta$. As $|\beta|\to \infty$,
any finite increment $\Delta$ is infinitismally small relative
to the magnitude of
$|\beta|$ and so we expect discrete derivative to yield
same result as  the usual derivative in this limit.
However, for finite $\beta$ such as equal to unity, we have
\be
\partial_{{\Delta},\beta} \tau(\beta)\vert_{\beta =1}
= {1\over (q-1)} \lim_{l\to 0} { {\rm ln}\;\sum_{i} (p_i)^{q} \over
 {\rm ln}\;l} =D_q,
\ee
which is the definition of generalized dimension.
This again illustrates that discrete derivative is the
generator of Renyi information, where using the ordinary derivative
would give Shannon information 
\be
{d\over d\beta}\tau(\beta)\bigg{\vert}_{\beta=1} = \lim_{l\to 0}\sum_i {p_i
 {\rm ln}\;p_i \over {\rm ln}\;l}.
\ee
\section*{4. Concluding remarks}
We have used discrete derivative to obtain generalized 
bit moments or cumulants and within that, we have 
noticed the  basis to distinguish Tsallis and Renyi 
entropy. Tsallis entropy has been previously related to
Jackson's $q$-derivative and this observation
motivated the search for connections between
quantum group theory and nonextensive statistics \cite{pla1}.
In fact, it is well known that discrete derivative
and Jackson derivative can be mapped onto each other \cite{dima}.
Briefly, the discrete translation produced by the former operator
correspond to dilatation  by the latter.
Thus the modification of the derivative in 
Eq. (\ref{subs}) can be written equivalently
as
 \be
 {\partial^k \over \partial\sigma^k}Z(\sigma) 
 \to (-yD_{y,q})^k Z(y),
 \label{rul}
 \ee
where $D_{y,q}$ is the  Jackson derivative operator
\be
D_{y,q}f(y) = {f(qy) - f(y)\over (q-1)y}.
\label{dd2}
\ee
 The variable $y$ is defined as $y = e^{-\sigma}$. Note that 
the operator $(-yD_{y,q})^k$ goes to ${\partial^k \over \partial\sigma^k}$
as $q\to 1$. 
 
Secondly, as both Shannon and Renyi entropies are additive with
respect to independent subsystems, and nonadditivity
has been thought to be related to Jackson derivative, it
seems Renyi entropy may not  have any connection with
discrete derivatives. This is somewhat surprising
as Jackson derivative seems relevant for
 multifractal sets \cite{erzjp} and Renyi entropy has also proved
useful for such systems \cite{becksh}. 
After showing above that discrete derivative 
is the generator of Renyi entropy, we point out  
 that it is also consistent with the additive property 
 of Renyi entropy which has been  
 derived here from the generating function $G = {\rm ln}\;Z$.
 Thus  for independent subsystems $A$ and $B$,
 $Z(A+B) = Z(A).Z(B)$ and so the generating function is 
additive $G(A+B) = G(A)+G(B)$. Now 
 the discrete derivative is distributive:
\be
\partial_{{\Delta},x}(f+g)(x)=\partial_{{\Delta},x}f(x)
          +\partial_{{\Delta},x}g(x).
\ee
 and additivity of Renyi entropy follows directly from this. 

In general, we observe that cumulants of method (i)
are nonadditive with respect to independent subsystems, whereas
those of method (ii) are additive in nature. 
Thus based on discrete derivative, for one class
of (nonextensive) $q$-bit moments in Eq. (\ref{mkq}), we have two 
classes of $q$-bit cumulants distinguished by their additive 
properties. Tsallis and Renyi entropies 
have been cast here in the more general framework of bit-moments and
cumulants.  Particularly, 
the concept of biased averaging $\sum_i (p_i)^q$
found in both the entropies 
is shown to have its origin in the finite increment
induced by discrete derivative
or equivalently, dilatation induced by Jackson derivative.  
 It is hoped that the present approach will help
in better understanding the origin and connection between  
these  entropies.  
\subsection*{Acknowledgements}
I wish to thank Alexander von Humboldt Foundation, Germany,
for financial support.

\end{document}